\documentclass[prb,twocolumn,amssymb,amsmath,nobibnotes]{revtex4}

\usepackage{graphicx,color}
\usepackage{subfigure}
\usepackage{dcolumn}
\usepackage{bm}
\usepackage{threeparttable}
\usepackage{stmaryrd}
\begin{document}

\newenvironment{figurehere}
  {\def\@captype{figure}}
  {}

\title{Strain effects on work functions of pristine and
       potassium-decorated carbon nanotubes}

\author{Yongqing Cai$^1$}
\author{Aihua Zhang$^1$}
\author{Yuan Ping Feng$^1$}
\email{hyfyp@nus.edu.sg}
\author{Chun Zhang$^{1,2}$}
\email{phyzc@nus.edu.sg;}
\affiliation{
	$^1$Department of Physics, National University of Singapore,
		2 Science Drive 3, Singapore 117542\\
	$^2$Department of Chemistry, National University of Singapore}

\author{Hao Fatt Teo}
\author{Ghim Wei Ho}
\affiliation{Engineering Science Programme, National University of Singapore, Singapore, 117574}

\date{\today}

\begin{abstract}
Strain dependence of electronic structures and work functions of both pristine and potassium doped (5,5) (armchair)
and (9,0) (zigzag) carbon nanotubes (CNTs) has been thoroughly studied using first-principles calculations based on
density functional theory (DFT). We found that for pristine cases, the uniaxial strain has strong effects on work
functions of CNTs, and the responses of work functions of CNT (5,5) and (9,0) to the strain are distinctly different.
When the strain changes from -10 \% to +10 \%, the work function of the CNT (5,5) increases monotonically from 3.95 eV
to 4.57 eV, and the work function of the (9,0) varies between 4.27 eV and 5.24 eV in a complicated manner. When coated
with potassium, for both CNTs, work functions can be lowered down by more than 2.0 eV, and the strain dependence of
work functions changes drastically. Our studies suggested that the combination of chemical coating and tuning of
strain may be a powerful tool for controlling work functions of CNTs, which in turn will be useful in future design of CNT-based electronic and field-emitting devices.

\end{abstract}

\maketitle

Carbon nanotubes (CNTs) have attracted tremendous interests since their discovery \cite{Iijima} in 1991. They have
been regarded as great candidates for building blocks of future nanoscale electronic and photonic devices due to their
remarkable electronic properties. In last two decades, large amounts of CNT-based devices such as field effect
transistors,\cite{Tans,Avouris} diodes,\cite{Harish} tunneling magnetoresistance junctions,\cite{ZC,GuoH}
and field emitters \cite{Bonard,Heer} have been proposed experimentally or theoretically. A complete understanding
 of electronic properties of CNTs and their dependence on various kinds of physical or chemical conditions holds
 the key for the future design and applications of these devices. In particular, several recent studies have
 shown that work functions of CNTs may be critically important in the functioning of some CNT-based devices \cite{Avouris,Harish,Zhao jijun_PRB,Bin Shan,W. S. Su}. For
 example, the Schottky barrier formed at metal-CNT junction that is closely related to work functions of two materials
 has been demonstrated to be crucial for performance of CNT-based field effect transistors and diodes.\cite{Avouris,Harish}
 It has also been shown by experiments that the efficiency of CNT-based field emitting devices sensitively depends
 on work functions of CNTs.\cite{Youngmi Cho} Therefore, a practical and effective method that can be used to modify, tune
 and control work functions of CNTs is highly desired.
\begin{figure}
  \includegraphics[width=9cm]{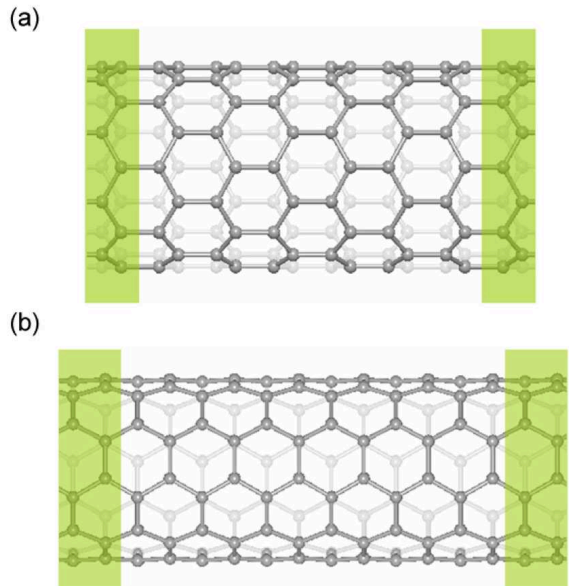}\\
  \caption{Supercells of  CNT(9,0) (upper) and CNT(5,5) (bottom) employed in calculations. Strains are applied by elongating or shortening the length of tubes along the tube axis (z axis) with two outermost layers (shadowed area) fixed to their bulk structures without strain. In all calculations, the periodically replicated CNTs (in both x and y directions) are separated by a vacumm region of 10 \AA. }\label{FIG.1}
\end{figure}

As a practical issue, strain in CNTs is important since mechanical deformations are inevitable in the process of
fabricating CNT-based devices. In literature, it has been pointed out that the structure deformation in a CNT often
drastically changes its intrinsic properties. Some experimental and theoretical studies showed that the conductance of
a metallic CNT could decrease by orders of magnitude when under a small strain around 5\%.\cite{Tombler, ZhangC_Strain}
Some other studies suggested that a metallic CNT may undergo a metal-semiconductor transition when strained, and the
bandgap of the CNT sensitively depends on the strain.\cite{S. Sreekala, Jun-Qiang Lu} Despite the importance of strain on electronic properties
of CNTs, detailed studies of strain effects on work functions of CNTs fell short.

In this paper, we present our theoretical investigations under the framework of density functional theory (DFT) on
correlations between uniaxial strain and work functions of pristine and potassium-decorated CNT (5,5) and CNT (9,0).
For pristine cases, we found that the strain has strong effects on work function for both CNTs, and CNT (5,5) responds
to the strain distinctly different from CNT (9,0) does. When coated with potassium, our calculations showed that the
strain dependence of work function for both CNTs changes drastically.

DFT calculations were performed with a plane wave basis with cutoff energy of 400 eV, and ultra-soft pseudo potentials \cite{Vanderbilt}
using VASP package.\cite{vasp} In all calculations, the generalized gradient approximation (GGA) in PW91 format
\cite{GGA_PW91} and 1$\times$1$\times$5 Monkhorst-Pack \textbf{k} sampling \cite{M-K} were used.
 Supercells for both CNTs employed in calculations are shown in Fig. 1. The uniaxial tensile (positive) or compressive (negative)
 strain is applied to CNTs by elongating (positive) or shortening (negative) CNTs along the tube axis followed by a
full relaxation of all atoms except for two outermost layers (as shown in Fig. 1) whose atomic configurations are
fixed to those without strain. The structural deformation caused by this kind of strain is nonuniform along the tube
axis, and mimics that in real CNT-based devices which often employ a metal-CNT-metal configuration and contain
nonuniform structure deformation due to the stretching or compressing by two metal contacts. The work function is
defined as the energy needed to remove one electron from the system, thus can be calculated by subtracting the
Fermi energy from the electrostatic potential in the middle of the vacuum.

\begin{figure}
  \includegraphics[width=9cm]{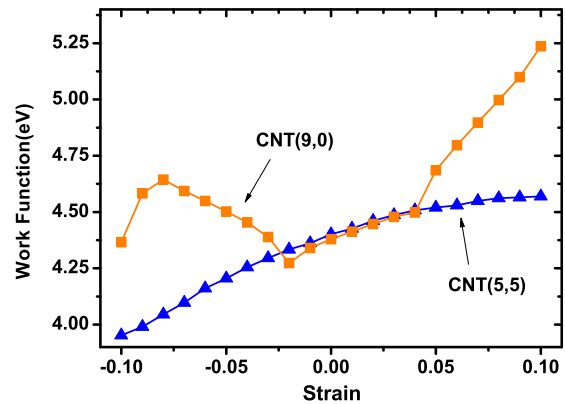}\\
  \caption{Work functions of CNT (5,5) and CNT(9,0) Vs. strain. Tensile strain is positive and compressive strain is negative. For CNT (5,5), the work function changes monotonically when strain changes from -10 \% to +10 \%. For CNT (9,0), when strain changes from -2\% and 4\%, the work function behaves similar to that of CNT (5,5).}\label{FIG.2}
\end{figure}

\begin{figure}
  \centering
 \includegraphics[width=8cm]{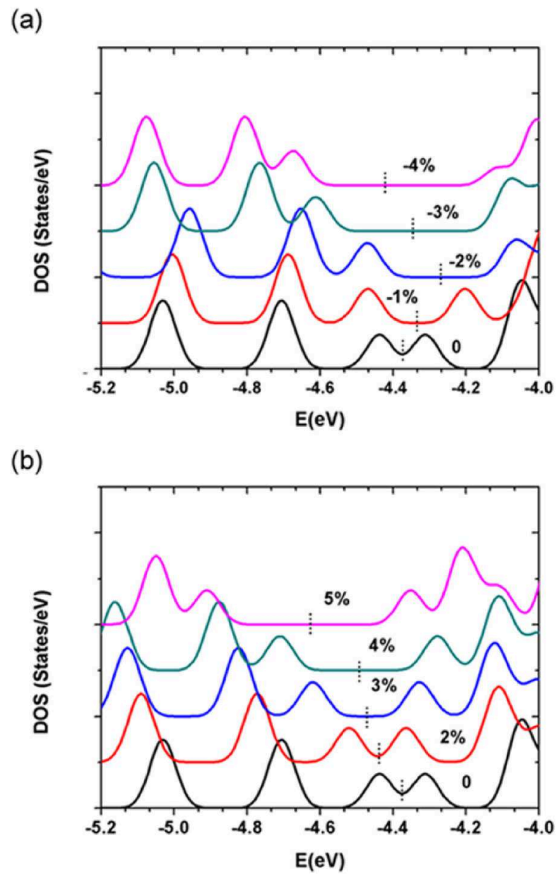}\\
  \caption{(Color online) Electron DOS as a function of energy for CNT(9,0) at different compressive (a), and tensile (b) strains. Dashed lines denote Fermi energies. Note: The electrostatic potential in the middle of vaccum is set to be 0.0 eV.}\label{FIG.3}
\end{figure}

Work functions of CNT(5,5) and CNT(9,0) without strain are estimated by our calculations to be 4.38 eV and 4.40 eV
respectively, which are consistent with previous studies.\cite{Zhao jijun_PRB,Youngmi Cho} We found that strain
has great effects on work functions of both CNTs, and when strained, the work function of CNT (5,5) behaves very
differently from that of CNT (9,0) as shown in Fig. 2. When strain changes from -10 \% to +10 \%, the work function
of the CNT (5,5) increases monotonically from 3.95 eV to 4.57 eV, whereas the work function of the CNT (9,0) varies between
4.27 eV and 5.24 eV in a complicated manner. Within this range of strain, the amount of changes of work function,
0.62 eV for CNT (5,5) and 0.97 eV for CNT (9,0), are quite significant, which shows strong effects of strain and
suggested that tuning strain may be an effective method to control the Schottky barrier at metal-CNT juncton. Since
previous studies have shown that the Schottky barrier at metal-CNT junction is critical for chemical bonding at
contacts,\cite{Vitale} and crucial for performance of CNT-based devices,\cite{Avouris, Harish} we believe that
understanding strain effects on work functions of CNTs is important for future design and control of nanoscale
CNT-based devices.

The complicated behavior of the work function of CNT(9,0) under strain is quite interesting. Within the range of strain
from -2\% to 4\%, the work function of CNT (9,0) behaves quite similar to that of CNT (5,5). When the strain is less
than -2\% or bigger than 4\%, the work function starts to significantly deviate from the monotonic behavior of
CNT (5,5). Here, the strain of -2\% and 4\% are two turning points. To understand this, we plotted the density of
states (DOS) of electrons for CNT (9,0) at compressive strain in Fig. 3 (a), and tensile strain in Fig. 3 (b). Since our calculations showed that the variations of electrostatic potential in the middle of the vacuum are very small when strain changes, in the figure, we set the reference energy (0.0 eV) to the electrostatic potential of vacuum. Several interesting things can be seen from this figure. First, when under strain, CNT (9,0)
undergoes a metal-semiconductor transition as predicted in literature.\cite{S. Sreekala}. Second, variations of work function under strain as shown in Fig. 2 can be completely understood by the changes of Fermi energy at different strains as depicted in Fig. 3. For example, for compressive strain, the Fermi energy increases from strain 0.0 to -2\%, and decreases from strain -2\% to 4\%, which leads to the decrease of work function from strain 0.0 to -2\%, and the increase of work function from strain -2\% to 4\% as shown in Fig. 2. The sudden decrease of Fermi energy from tensile strain 4\% to 5\% corresponds to the sudden increase of work function at the same tensile strain as shown in Fig. 2. It is worthy mentioning here that the Fermi energy is calculated from the charge neutrality which is actually determined by the tail of DOS of HOMO-like states. Therefore, underlying the change of Fermi energy is the change of HOMO states due to strain which in turn determines the change of work function. We also calculated DOS for CNT(5,5) under different strains (not shown in this paper). For CNT (5,5), there is no metal-semiconductor transition that agrees with literature,\cite{L Yang,Amitesh Maiti} and DOS changes continuously with strain for all cases. The electronic structures of CNT (9,0) and CNT (5,5) are intrinsically different:
(9,0) is a semi-metal which will undergo a metal-semiconductor transition under a small strain, while (5,5) is truly metallic, and this is the reason
why their work functions respond completely differently to strain. We believe that our calculations shown here for CNT (5,5) and (9,0) are helpful for
understanding strain effects on work functions of other CNTs.

\begin{figure}
  \includegraphics[width=9cm]{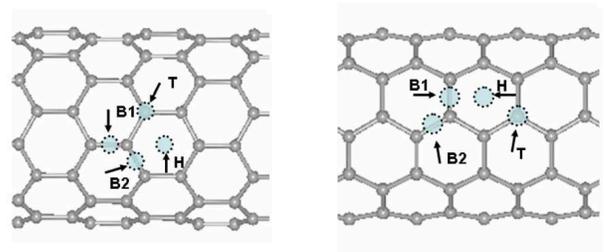}\\
  \caption{A schematic description of different adsorbing sites(purple circles) of K atom on CNT(9,0) (left) and CNT(5,5) (right): Bridge sites \emph{B1}, \emph{B2}; Hollow site \emph{H}, and top site \emph{T}. Binding energies for different sites are calculated to be \emph{B1}: 1.58 eV, \emph{B2}: 1.62 eV, \emph{H}: 1.73 eV, \emph{T}: 1.59 eV for CNT(9,0), and \emph{B1}: 0.95 eV, \emph{B2}: 0.92 eV, \emph{H}: 1.04 eV, \emph{T}: 0.92 eV for CNT(5,5),repectively. }\label{FIG.4}
\end{figure}

\begin{figure}

  \includegraphics[width=9cm]{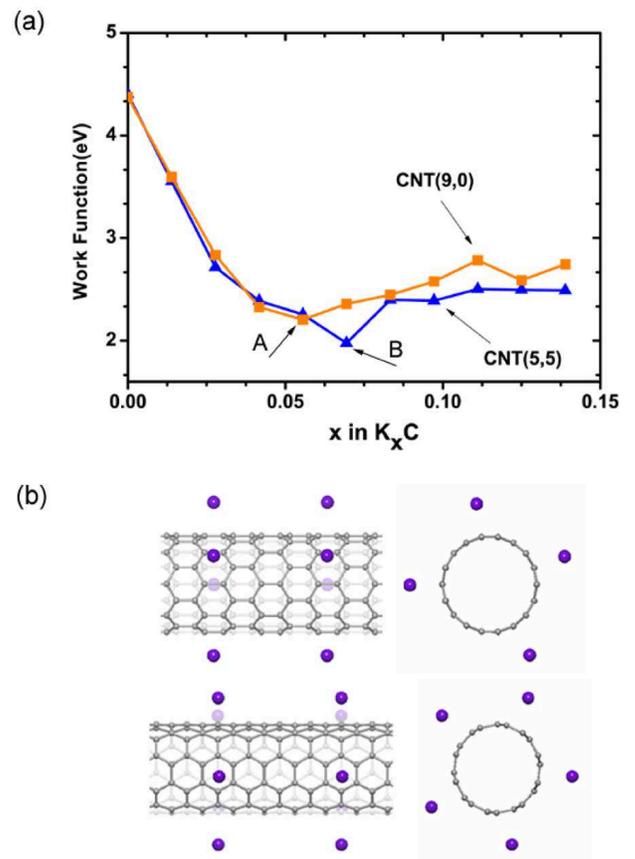}
  \caption{Work function of potassium deposited CNT (9,0) and CNT (5,5) Vs. deposition density(x in KxC).
    For CNT (9,0), the work function reaches its minimum value of 2.2 eV at the coating density of 5.56\%
  (denoted as A in (a)), and the minimum work function of CNT (5,5) is 1.98 eV occurring at the coating
  density of 6.25\% (B in (a)). The atomic structures of K-coated CNT (9,0)  and (5,5) with minimum work
  functions are shown in Fig. 5 (b) ( A: upper, and B: lower).}
  \label{FIG. 5}

\end{figure}

\begin{figure}
  \includegraphics[width=9cm]{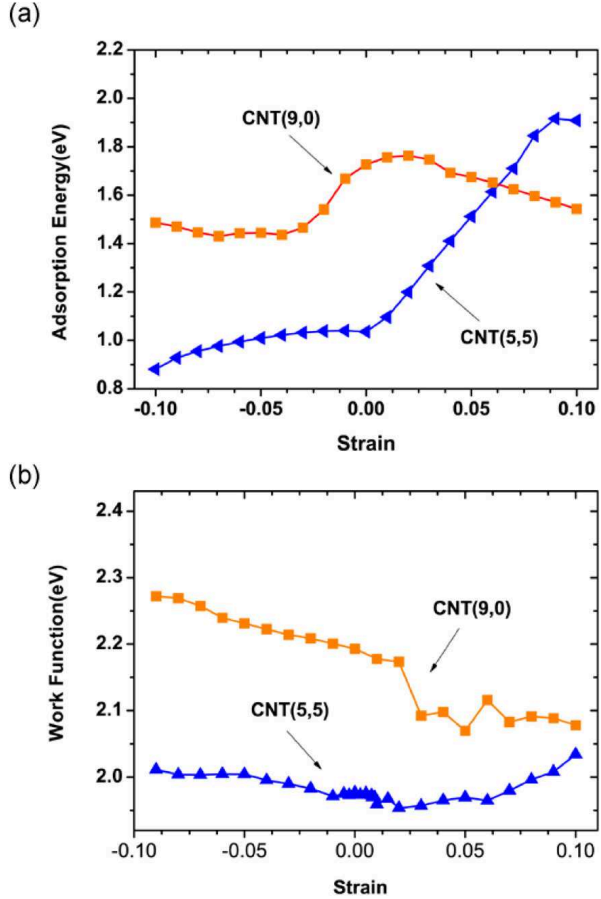}\\
  \caption{Strain effects on adsorption energies of K@CNTs, and work functions of CNTs: (a) Adsorption energy of a single K atom on CNT (9,0) and (5,5) as a function of strain; (b) Strain dependence of work functions of K coated CNT (9,0) as shown in Fig. 5 (b) (upper), and K coated CNT (5,5) (lower one in Fig. 5 (b)).}\label{Fig.6}
\end{figure}

Next, we examine the strain effects on work functions of potassium-decorated carbon nanotubes. It has been well-known
that the coating of alkali metals can greatly reduce the work function of CNTs, which may significantly
improve the efficiency of CNT-based field-emitting devices.\cite{G. Zhao} It is naturally interesting to see if we can gain further
control of the work functions of CNTs by combining alkali metal coating and tuning strain. First, we test the effects
of potassium coating on CNT (9,0) and (5,5) without strain. For this purpose, we calculated the adsorption energies
of a single K atom on different adsorbing sites of CNT (9,0)  and CNT (5,5) (as shown in Fig. 4), and found that for both CNTs,
the most stable adsorption site is the hollow site with the adsorption energy of  1.73 eV for CNT (9,0) and 1.04 eV for
CNT (5,5), agreeing well with literature \cite{E. Durgun}. We then gradually increased the number of K atoms adsorbed per supercell
on CNTs, and then calculated the change of work functions after the structure optimization. As shown in Fig. 5, for both CNTs, when
the coating density increases, the work function first rapidly decreases by more than 2.0 eV, and then slowly goes up.
For CNT (9,0), the work function reaches its minimum value of 2.2 eV at the coating density of 5.56\% (denoted as A in
Fig. 5 (a)), and the minimum work function of CNT (5,5) is 1.98 eV occurring at the coating density of 6.25\% (B in
Fig. 5 (a)). The atomic structures of K-coated CNT (9,0)  and (5,5) with minimum work functions are shown in
Fig. 5 (b) ( A: upper, and B: lower). Our calculations are in good agreement with observations of previous
studies that a great reduction of work functions of CNTs can be achieved by alkali metal coating.\cite{G. Zhao,Satoru Suzuki}

\begin{figure}
  \includegraphics[width=8
  cm]{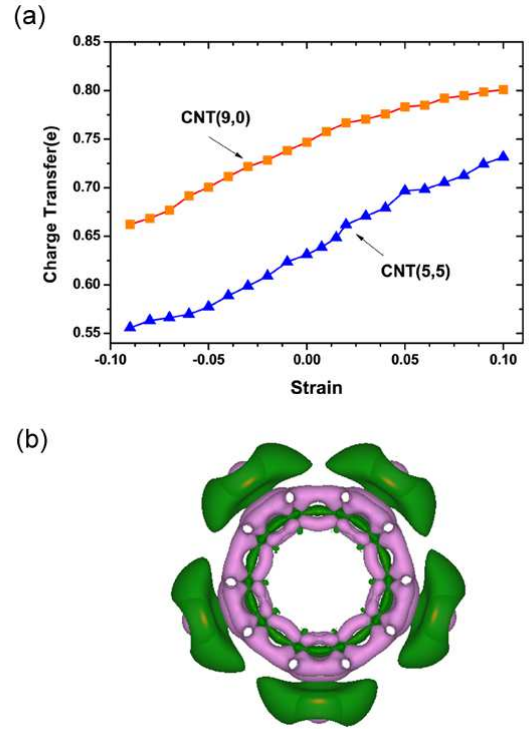}\\
  \caption{(a) Calculated average charge transfer per K atom to CNT(5,5) and CNT(9,0) at different strains. When strain changes from -10\% to 10\%, the charge transfer for two tubes monotonically increases. (b) Isosurface of the differential charge density between isolated CNT (5,5), K atoms, and
K-coated CNT (5,5).}\label{Fig.7}
\end{figure}

The uniaxial strain as described before is then applied to K-coated CNT (5,5) and CNT (9,0). Before discussing the
change of work functions, we first take a look at effects of strain on adsorption energies of a single K atom on
both CNTs. As shown in Fig. 6(a), the strain has much stronger effects on adsorption energy of K on CNT (5,5) than that
of CNT (9,0) especially for tensile strain. From 0.0 to 10\%, the adsorption energy of K on CNT (5,5) increased by as
much as 0.87 eV. The significant change of adsorption energy under strain implies potential applications of modifying
chemical properties of CNT-based systems via tuning strain. The work functions Vs. strain were calculated for two
K-coated CNTs (as shown in Fig. 6 (b)) which have lowest work functions upon K coating. Interestingly, when under
strain, work functions of K-coated CNTs show drastically different behaviors comparing with those of pristine cases.
First, the strain dependence becomes much weaker, and second,
work functions decrease when strain increases except for the case of K-coated CNT (5,5) in the small range of strain from
 6\% to 10\%. When strain changes from -10\% to
+10\%, the work function of K-coated CNT (9,0) varies between 2.27 eV and 2.08 eV, and the work function of K-coated
CNT (5,5) changes between 2.05 eV and 1.95 eV for all strains. These drastic changes of strain-dependence of work functions upon K coating strongly suggest that the combination of the chemical decoration and tuning of strain is a powerful method in controlling work functions of CNT-based systems.

In order to understand the decrease of work functions of K-coated CNTs, charge transfer between K
atoms and two CNTs as function of strain was calculated via Bader charge analysis.\cite{G. Henkelman} In Fig. 7 (a), we plot the calculated charge transfer per K atom for the K-coated tubes under strain ranging from -10\% to +10\%. For the case of zero strain, the amount of charge transfer per K atom was found to be 0.63 and 0.75 electrons for CNT(5,5) and CNT(9,0) respectively. In
Fig. 7 (b), we show the isosurface of the differential charge density between isolated CNT (5,5), K atoms, and
K-coated CNT (5,5). The plum (green) color denotes the accumulation (diminishing) of electrons. For both tubes,
the amount of charge transferred from K atoms to CNTs monotonically increases with strain. Therefore, it can be
expected that the dipole moment normal to tube surfaces caused by the charge transfer increases with strain, which will likely lead to the decrease of work function according to previous studies.\cite{U. Martinez}

In summary, variations of work functions with strain for both pristine and potassium coated CNT(5,5) and CNT(9,0)
are calculated via first principles methods under the framework of DFT. For pristine cases, the strain shows strong
effects on work functions of both CNTs, and the work function of CNT (5,5) behaves quite differently with strain
from that of CNT (9,0). The different strain dependence of work functions of two CNTs is caused by intrinsic
difference between metallic CNT (5,5) and semi-metallic CNT (9,0). We also found that the strain has great influence
on chemical adsorption of K atoms on CNTs. In the case of CNT (5,5), the adsorption energy of a single K atom increases by 0.87 eV when strain changes from 0.0 to 10\%.
At last, work functions of K-coated CNTs show drastically different strain dependence from that pristine cases.
Our findings strongly suggest that tuning of strain may be a powerful method in controlling work functions of CNT-based systems,
and our calculations on CNT (5,5) and CNT (9,0) are helpful for understanding work functions of other CNTs under strain.

Acknowledgement: This work is supported by ...

\end{document}